\documentclass[twocolumn,twoside,slac_two]{revtex4}
\usepackage{graphicx}
\usepackage{fancyhdr}
\begin{document}

\title{Leptonic and Hadronic Models for the Extra Components in
Fermi-LAT GRBs}

%

\author{K. Asano}
\affiliation{Interactive Research Center of Science, 
Tokyo Institute of Technology, 2-12-1 Ookayama, Meguro-ku, Tokyo 152-8550, Japan}
\author{P. M\'esz\'aros}
\affiliation{Dept. of Astronomy \& Astrophysics; Dept. of Physics; Center for Particle Astrophysics, Pennsylvania State University, University Park, PA 16802, USA}
\author{K. Murase}
\affiliation{Dept. of Physics; Center for Cosmology and AstroParticle Physics, Ohio State University, 191 West Woodruff Avenue, Columbus, OH 43210, USA}
\author{S. Inoue, T. Terasawa}
\affiliation{Institute for Cosmic Ray Research, University of Tokyo, Kashiwa-no-ha 5-1-5, Kashiwa-shi, Chiba 277-8582, Japan}

\begin{abstract}
{\it Fermi} satellite has detected extra spectral components in GeV energy range in
several GRBs. Those components have power-law shapes, which may contribute
to also X-ray band. The limited photon statistics make it difficult to determine the
origin of GeV photons, namely internal or external shocks. We try to explain the
extra components with our numerical simulations based on internal dissipation
picture. Our leptonic model may reproduce not only the GeV excess
via SSC emission but also the low-energy excess via the late synchrotron emission
from remnant electrons.
The hadronic models also reproduce keV-GeV power-law components
by synchrotron and SSC emissions from secondary electron-positron pairs.
In most cases the hadronic models
require a much larger energy of protons than gamma-rays.
However, the keV-GeV flat spectra detected in GRB 090902B
is well explained with a comparable energy in protons and gamma-rays
Finally, we discuss both advantages and weaknesses for both the leptonic
and hadronic models.
To overcome difficulties in internal dissipation models,
we propose introduction of
continuous acceleration similar to the second-order Fermi acceleration.
\end{abstract}

\maketitle

\thispagestyle{fancy}


\section{EXTRA-SPECTRAL COMPONENTS}

Most of the prompt emission spectra of gamma-ray bursts (GRBs)
can be described by the well-known Band function \citep{ban93};
the photon number spectrum
$\propto \varepsilon^{\alpha}$
below $\varepsilon_{\rm p}$,
and $\propto \varepsilon^{\beta}$ above it.
The spectral peak energy $\varepsilon_{\rm p}$
are usually seen in the MeV range.
One of main scientific targets for the {\it Fermi} satellite
was to investigate whether the spectral shape is consistent
with the Band function even in the GeV band.
The {\it Fermi} detected GeV photons from
several very bright bursts ($E_{\rm iso}>10^{54}$ erg) such as
GRB 080916C \citep{916C}, GRB 090902B \citep{902B}, and
GRB 090926A \citep{926A}.
In such bursts the onset of the GeV emission is delayed with respect to 
the MeV emission.
Some of them also have an extra spectral component above a
few GeV, distinct and additional to the usual Band function.
Interestingly, GRB 090902B and GRB 090510 \citep{abd09b,ack10} show
a further, soft excess feature below $\sim 20$ keV,
which is consistent with a continuation of the GeV power-law component.
While such spectra may be explained by the early onset of the afterglow
\citep{ghi10,kum10}, here we pursue possibility of internal-shock origins.

\section{HADRONIC MODELS}

\begin{figure*}[t]
\includegraphics[width=70mm]{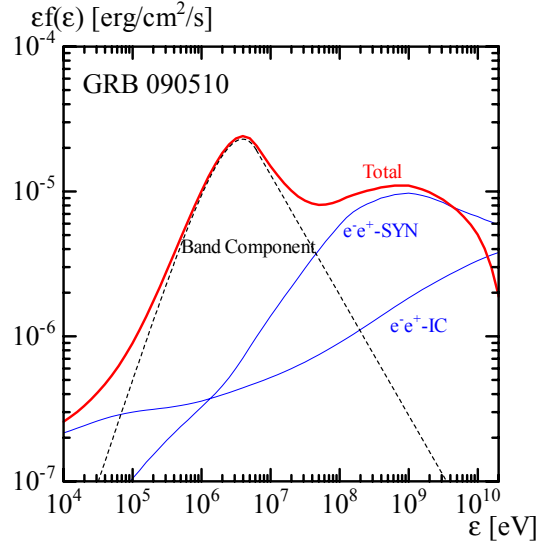}
\caption{Simulated spectra with hadronic models for GRB 090510.
The assumed fraction of the magnetic energy density
to the Band component photons is $10^{-3}$,
and the required proton luminosity
is 200 times the luminosity of the Band component.}
\label{f1}
\end{figure*}

If the spectral excesses in GeV and keV bands have
the same origin, such a wide photon-energy range may imply
the cascade emission due to hadrons.
If the GRB accelerated ultra-high-energy protons,
synchrotron and inverse Compton (IC) emission from
an electron-positron pair cascade
triggered by photopion interactions of
the protons with low-energy photons \citep{boe98,gup07,asa07}
can reproduce power-law photon spectra
as seen in {\it Fermi}-LAT GRBs.
Through Monte Carlo simulations,
Asano et al. \citep{asa09b} have shown that
a proton luminosity much larger than gamma-ray luminosity is required
to produce the extra spectral component in GRB 090510
as $L_{\rm p}>10^{55}$ erg $\mbox{s}^{-1}$ (see Fig.\ref{f1}).
Namely, the efficiency of photopion production is very low.
In this case, the spectrum of the GeV component
is very hard with photon index $\sim -1.6$,
which requires a inverse Compton (IC) contribution from the secondary pairs.
The prominent IC component leads to
a weaker magnetic field.
This entails a lower maximum energy of protons,
and hence lower photopion production efficiency.
Therefore, the required proton luminosity
is so large in GRB 090510.

\begin{figure*}[t]
\includegraphics[width=70mm]{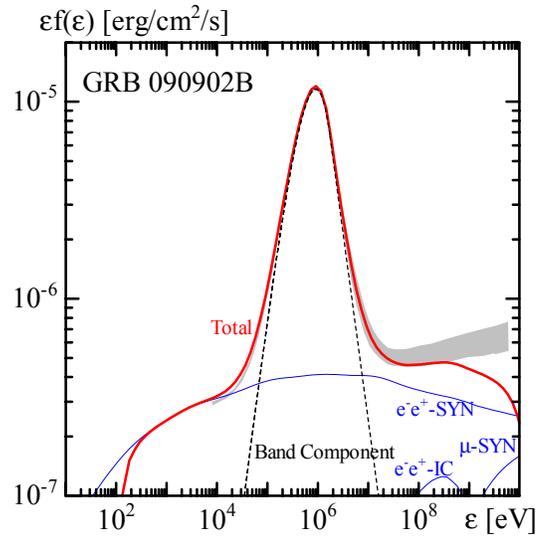}
\caption{Simulated spectra with hadronic models for GRB 090902B.
The assumed fraction of the magnetic energy density
to the Band component photons is $1.0$, and
the required proton luminosity
is 3 times the luminosities of the Band component.}
\label{f2}
\end{figure*}

The assumed bulk Lorentz factor in Fig.\ref{f1} is 1500,
and the emission radius is $R=10^{14}$ cm.
If we adopt a smaller value of $\Gamma$,
the pion production efficiency would increases as
$t_{\rm exp}/t_\pi \propto R^{-1} \Gamma^{-2}$.
However, we should note that there is a lower limit to $\Gamma$,
which is required to
make the source optically thin to GeV photons.
Given the photon luminosity and spectral shape,
this minimum Lorentz factor can be estimated as shown
in the online supporting materials
in Abdo et al. (2009) \citep{916C}.
In order to lower $\Gamma$, we have to increase
the emission radius $R$.
The lower limit of the Lorentz factor
$\propto R^{1/(\beta-3)}$ does not decrease drastically
(since $\beta \simeq -3$, $\Gamma_{\rm min} \propto R^{-1/6}$).
The required large luminosity is rather favorable
for the GRB-UHECR scenario, but it imposes stringent requirements
on the energy budget of the central engine.

On the other hand, GRB 090902B is encouraging for the hadronic model
because of its flat spectrum (photon index $\sim -2$)
\citep{asa10}.
The Band component in this burst
is an atypically narrow energy distribution
as shown in Fig. \ref{f2},
which may imply the photospheric emission \citep{ryd10}.
The hadronic cascade emission can well reproduce the
observed flat spectra including
the soft excess feature below 50 keV
(model parameters: $R=10^{14}$ cm, $\Gamma=1300$).
Owing to the flat spectral shape of the extra component,
no IC component is required.
We can adopt a strong magnetic field,
which enhance the photomeson production efficiency.
In this case the flux of the extra component
is relatively low compared to the Band component,
which also decreases the required proton luminosity.
Therefore, the necessary nonthermal proton luminosity is
then not excessive and only comparable to the Band component luminosity.

\begin{figure*}[t]
\centering
\includegraphics[width=65mm]{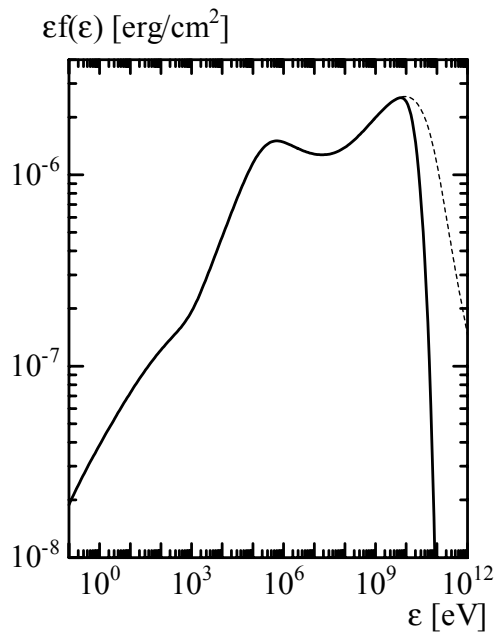}
\caption{The SSC-model fluence obtained from our
time-dependent simulation (assuming $z=1$).
The dashed line neglects absorption due to EBL.} \label{f3}
\end{figure*}

\section{LEPTONIC MODELS}

As Corsi et al. \citep{cor10} discussed, the GeV emission may be due to IC
emission from internal dissipation regions.
However, it seems difficult to explain spectral excesses
in both keV and GeV bands by IC emission.
Recently, we carry out time-dependent simulations
of photon emissions with leptonic models \citep{asa11}.
In our simulations,
as the photon energy density increases with time
because of synchrotron emission,
the SSC component gradually grows and dominate
the photon field later.
This late growth of the IC component has been
observed also in the simulations of Bo\v{s}njak et al. (2009)
\citep{bos09}.
The resultant lightcurves show delayed onset
of GeV emission, but the delay timescale would be within 
the approximate timescale of the keV-MeV pulse width.
However, the longer delay compared to the pulse 
timescale such as observed in GRB 080916C
is not explained by this effect only.

As shown in Fig. \ref{f3}
(model parameters: $R=6 \times 10^{15}$ cm,
$\Gamma=1000$, $B=100$ G,
$E_{\rm iso}=10^{54}$ erg,
$\varepsilon_{\rm e,min}=11.3$ GeV), the model spectrum
obtained from our simulations reproduce
both the low and high energy excesses.
When the magnetic field is weak enough,
even at the end of the electron injection,
the cooled electrons can be still relativistic.
Such cooled electrons
continue emitting synchrotron photons.
The cooling due to IC gradually becomes inefficient as
the seed photon density decreases.
Such late synchrotron emission can yield
the low-energy excess, while IC emission
makes a GeV extra component.

\begin{figure*}[t]
\centering
\includegraphics[width=80mm]{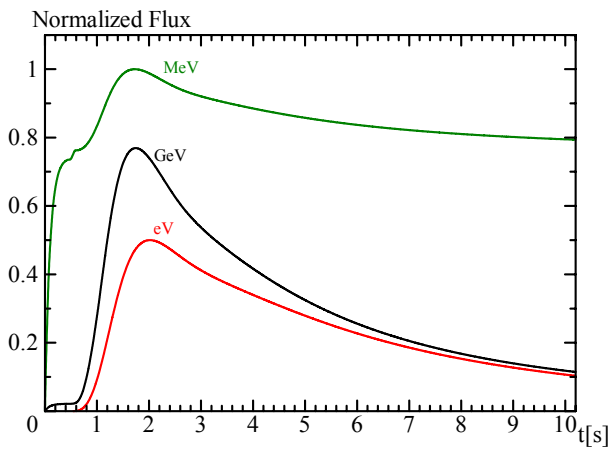}
\caption{The lightcurves for the external IC model.
In the MeV band the external photons from inside region dominate,
while GeV and eV emissions are orginated from accelerated electrons
in outside dissipation region.} \label{f4}
\end{figure*}

Another possible model is the external IC emissions \citep{tom09,tom10},
which can explain the delayed onset of GeV emission.
The spatial separation between the source of the external photons
and the site of the internal shock region corresponds
to the delay timescale.
We also carry out a simulation for this model
assuming an external emission $L_{\rm seed}=10^{53}~{\rm erg}~{\rm s}^{-1}$
with the Band function:
$\varepsilon_{\rm peak}\simeq 1$ MeV in the observer frame,
$\alpha=-0.6$, and $\beta=-2.6$.
The parameters for the internal shock
are similar to those in Fig.\ref{f3} as $R=6 \times 10^{15}$ cm,
$\Gamma=1000$, $B=100$ G,
$E_{\rm iso}=3 \times 10^{53}$ erg, except for
$\varepsilon_{\rm e,min}=50.6$ MeV.
Our simulation reproduces the delayed onset of GeV emission,
GeV extra component, and softening of the Band component.
Since the external photon field in the rest frame of the emission region
is highly anisotropic, the marginally high-latitude emission contributes
the most to the flux.
Thus, the simulated GeV lightcurve shows larger delayed onset
and longer tail than that in the usual leptonic models (Fig.\ref{f4}).

\section{IMPLICATION}

The hadronic models usually require a much larger
energy of protons than observed gamma-rays
except for some examples such as GRB 090902B.
Some may consider
that the leptonic SSC models seem rather reasonable to
reproduce both the GeV and keV excesses.
However, a lower magnetic field and high Lorentz factor,
required to make an optically thin source for GeV
photons generated via IC emission,
leads to a very high minimum Lorentz factor ($\gamma_{\rm e,min}
\sim 10^4$)
for random motion of accelerated electrons.
If all electrons are accelerated in internal shock regions,
such a high value may be impossible.
Thus, the fraction of accelerated electrons should be
small in leptonic models to explain {\it Fermi}-LAT GRBs.

On the contrary, the minimum Lorentz factor of
electrons $\gamma_{\rm e,min}$
in the external IC model should be small
to adjust the energy of scattered photons.
Given the total energy, such small $\gamma_{\rm e,min}$
means large number of electrons.
If the numbers of electrons and protons are the same,
the proton energy becomes fairly large
($\sim 1.9 \times 10^{54}$ erg in the case of Fig.\ref{f4},
while the luminosity of the Band component
$L_{\rm seed}=10^{53}~{\rm erg}~{\rm s}^{-1}$).
To enhance the emission efficiency,
electron-positron plasma should be introduced in such models.

Despite the ability of our straightforward simulations to reproduce various observed
properties of the GRB prompt emission,
the discrepancy with the low-energy index $\alpha$ remains unexplained.
The injected electrons cool via synchrotron radiation,
and distribute below $\gamma_{\rm e,min}$ with
a power-law index $-2$.
The resultant photon index becomes $\alpha \sim -1.5$,
while the observed typical values are $-1.0$ or harder.
The low-energy photon index affects
the photomeson production efficiency.
We have also tried to resolve this problem \citep{asa09c,mur11}
considering continuous acceleration/heating due to magnetic turbulences
induced via various types of instabilities
such as Richtmyer-Meshkov instability \citep{ino11}.
The acceleration/heating balances with the synchrotron cooling,
so the observed low-energy spectral index is naturally explained.
Such effect should be included in the time-dependent code
to reproduce the global shape of the GRB prompt spectra
from eV to GeV.
Especially, electron injection due to hadronic processes
and succeeding acceleration by turbulences may
explain very high $\gamma_{\rm e,min}$ and
GeV emission.

We plan to develop the time-dependent code
shown here to treat hadronic processes or
continuous acceleration/heating.
Note that the results for the hadronic models shown here
were calculated based on the steady state approximation.
We will carry out simulations for 
various situations involving dissipative photospheres
and internal or external dissipation or shock regions.
Moreover, the code will be useful for simulating emissions
of other high-energy sources, such as active galactic nuclei,
supernova remnant, and clusters of galaxies.

\begin{acknowledgments}
The series of our studies introduced here are
partially supported by Grants-in-Aid for Scientific Research
No.22740117 (KA), No.22540278 (SI),
and No.21540259 (TT) from the Ministry of Education,
Culture, Sports, Science and Technology (MEXT) of Japan.
\end{acknowledgments}

\bigskip 


\begin{thebibliography}{9}   


\bibitem{ban93}
Band, D. et al. 1993, ApJ, 413, 281
\bibitem{916C}
Abdo, A. A. et al., 2009, Science, 323, 1688
\bibitem{902B}
Abdo, A. A. et al. 2009, ApJ, 706 L138
\bibitem{926A}
Ackermann, M. et al.,  2011, ApJ, 729, 114
\bibitem{abd09b}
Abdo, A. A. et al. 2009, Nature, 462, 331
\bibitem{ack10}
Ackermann, M. et al. 2010, ApJ, 716, 1178
\bibitem{ghi10}
Ghisellini, G. at al. 2010, MNRAS, 403, 926
\bibitem{kum10}
Kumar, P., \& Barniol Duran, R. 2010, MNRAS, 409, 226
\bibitem{boe98}
B\"ottcher, M., \& Dermer, C. D. 1998, ApJ, 499, L131
\bibitem{gup07}
Gupta, N., \& Zhang, B., 2007, MNRAS, 380, 78
\bibitem{asa07}
Asano, K., \& Inoue, S. 2007, ApJ, 671, 645
\bibitem{asa09b}
Asano, K., Guiriec, S., \& M\'esz\'aros, P. 2009, ApJ, 705  L191
\bibitem{asa10}
Asano, K., Inoue, S., \& M\'esz\'aros, P. 2010, ApJ, 725, L121
\bibitem{ryd10}
Ryde, F. et al. 2010, ApJ, 709, L172
\bibitem{cor10}
Corsi, A. Guetta, D., \& Piro, L.  2010, A\&A, 524, 92
\bibitem{asa11}
Asano, K., \& M\'esz\'aros, P. 2011, ApJ, 739, 103
\bibitem{bos09}
Bo\v{s}njak, \v{Z}, Daigne, F., \& Dubus, G. 2009, A\&A, 498, 677
\bibitem{tom09}
Toma, K., Wu, X.-F. \& M\'esz\'aros, P. 2009, ApJ, 707, 1404
\bibitem{tom10}
Toma, K., Wu, X.-F. \& M\'esz\'aros, P. 2010, MNRAS, 415, 1663
\bibitem{asa09c}
Asano, K., \& Terasawa, S. 2009, ApJ, 705, 1714
\bibitem{mur11}
Murase, K., Asano, K., Terasawa, S., \& M\'esz\'aros, P.
2011, arXiv:1107.5575
\bibitem{ino11}
Inoue, T., Asano, K., \& Ioka, K. 2011, ApJ, 734, 77

\end{thebibliography}
\end{document}